# Ab initio study of nonlinear optical susceptibilities in silicon nanowires

**Daryoush Shiri** *

Department of Physics, Chalmers University of Technology, SE-412 96 Göteborg, Sweden

* Corresponding author: e-mail shiri@chalmers.se

Using Time Independent Density Functional Theory (TIDFT) it is shown that the 2$^{nd}$ order optical susceptibilities of narrow (1nm-2nm) Silicon Nanowires (SiNW) are enhanced due to surface termination. The value of $\chi^{(2)}$ is enhanced up to 200 pm/V which is promising a strong Second Harmonic Generation (SHG) in SiNWs. For [100], [110] and [111] SiNWs, *yxx* component of $\chi^{(2)}$ tensor is 81, 225 and 81 pm/V, respectively. These are in close agreement with $\chi^{(2)}$ values reported for strained silicon waveguides in experiments. The 3$^{rd}$ order susceptibility, $\chi^{(3)}$, is within the range of $(0.1\text{-}12)\times 10^{-18}$ m$^2$/V$^2$ which is close to the experimental values of bulk silicon $(0.1\text{-}0.2)\times 10^{-18}$ m$^2$/V$^2$ for [110] and [100] SiNWs and it is 100 times better for [111] SiNW. This study suggests possibilities of enhancing SHG in SiNWs through symmetry breaking via strain and surface termination/reconstruction as well as suitability of this DFT-based method in predicting nonlinear optical susceptibilities of nano structures.

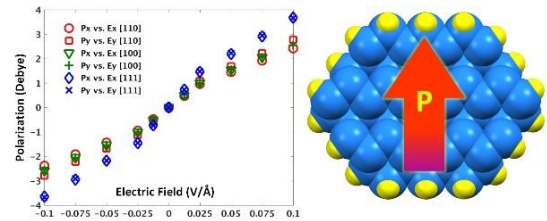

**1 Introduction** Centro-symmetric crystal of bulk silicon leads to zero dipolar second order nonlinear optical susceptibility [1, 2]. As a quest to use silicon for active optoelectronic devices, there has been intensive research on different methods to break the centro-symmetricity. Abrupt termination of silicon surface with SiO$_2$ [3, 4], interfacing with silicon nitride (and the induced strain due to that [5, 6]), and applying electric field [7] are among the methods with which $\chi^{(2)}$ is enhanced. For example the strain due to Si$_3$N$_4$ top layer in silicon nanowire waveguides leads to $\chi^{(2)}$ = 800 pm/V and 240 pm/V in electro-optical [5] and all optical devices [6], respectively. Motivated by this, the effect of surface termination and residual stress in breaking the symmetry of narrow (1nm-2nm) SiNWs and inducing second order optical nonlinearity was investigated. Although Time Dependent DFT-based methods are available to calculate optical susceptibilities [8, 9], the TIDFT method implemented in SIESTA® was adopted due to its linear scale (order N) and simplicity in obtaining reliable first order approximation for susceptibilities [10, 11]. What is common in TDDFT and TIDFT is that they both work based on perturbing the system by an external electric field (**E**) and calculating the induced dipole polarization (**P**). Hence the applied electric field should have zero component along the periodic direction of the solid e.g. along the length of the nanowire. For this reason some components of susceptibility tensors are not calculable. In this article it is shown how symmetry breaking by surface termination and residual stress in SiNWs lead to enhanced 2$^{nd}$ order nonlinearity. The calculated $\chi^{(2)}$ values are in the same ballpark of the values reported in [5] and [6] in which the strain induced due to SiO$_2$ substrate increases $\chi^{(2)}$ from 15 pm/V to more than 60 pm/V.

The calculated $\chi^{(3)}$ values closely match with that of bulk silicon and they are even 100 times better than bulk for [111] SiNWs. Vital for the application in solar cells, the Two Photon Absorption (TPA) coefficient can be estimated from $\chi^{(3)}$ values and it is shown that they are within the range of 10-20 cm/GW. This is in close agreement with the values calculated by Iitaka et al. for silicon quantum dots [12].

The rest of this article is organized by starting from the methods which include energy minimization and calculation of polarization in response to the applied static electric field. This is followed by presenting and discussing the results. Prospects for new applications, advantages and shortcomings of the method conclude the article.

## 2 Methods

**2.1 Structure and Relaxation** The nanowires in this work are cut from bulk silicon along [110], [100] and [111] crystallographic directions. The average diameters are 1.7 nm, 1.1 nm and 0.6 nm, respectively. The energy minimization is performed with DFT method of SIESTA® package using exchange-correlation functional of Generalized Gradient Approximation (GGA) type with Perdew-Burke-Ernzerhof (PBE) pseudopotentials [10]. The number of k-point samples within the Brillouin Zone (BZ) is 1×1×40 with highest number of points along the axis of the nanowire ($z$ axis). The minimum distance of adjacent unit cells is 60 Å to avoid interaction due to wave function overlapping. Energy cut-off, split norm, and force tolerance are 680 eV, 0.15, and 0.01 eV/A°, respectively. The energy of the unit cell of SiNW is minimized using Conjugate Gradient (CG) algorithm with variable unit cell option. Figure 1. shows the cross section of a Hydrogen passivated 1.7nm SiNW in [110] direction. The cross sections of a 1.1 nm [100] and a 0.6 nm [111] SiNW are shown in Fig. 2.

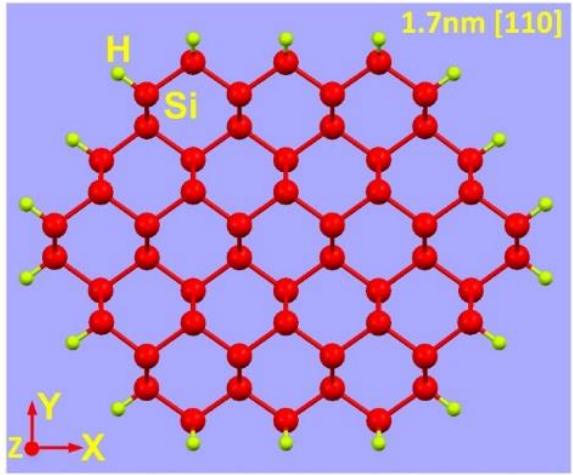

**Figure 1** Cross section of a 1.7 nm [110] SiNW. Dark and bright atoms are silicon and hydrogen, respectively.

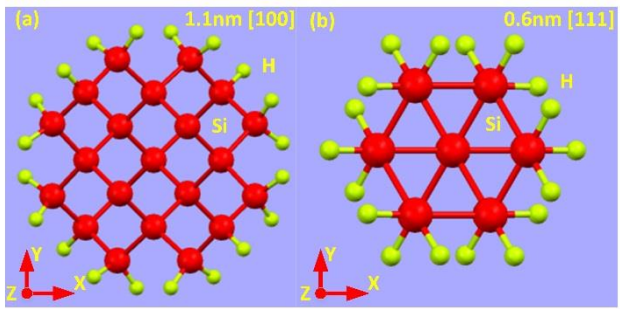

**Figure 2** Cross sections of (a) 1.1 nm [100] and (b) 0.6 nm [111] SiNWs. Dark and bright atoms are silicon and hydrogen, respectively.

**2.2 Nonlinear Susceptibility** After energy minimization, the unit cell of the SiNW is clamped and an external electric field (**E**) is applied in parallel with $x$ or $y$ directions. The electric field is varied within [-0.1,+0.1]V/A° range and for each given value, three components of Polarization (**P**) are calculated using SIESTA® [10, 11]. Generally the tensor relation between P and E is written as [1]:

$$\vec{P} = P_0 + \epsilon_0 \chi^{(1)} \vec{E} + \epsilon_0 \chi^{(2)} \vec{E}.\vec{E} + \epsilon_0 \chi^{(3)} \vec{E}.\vec{E}.\vec{E} \quad (1)$$

where $P_0$ is static polarization (under zero electric field), $\chi^{(1)}$, $\chi^{(2)}$ and $\chi^{(3)}$ are second, third and fourth rank tensors, respectively i.e. each of which has 9, 27 and 81 components. For SiNWs the $z$ component of electric field ($E_z$) should be zero i.e. $E_z=0$ as $z$ is the periodic direction of the nanowires. Hence the electric field is within the $xy$ cross sectional plane and its $x$ and $y$ components are changed to calculate $P_x$, $P_y$ and $P_z$. As an example equation (1) is expanded to derive the $y$ component of polarization keeping in mind that $E_z=0$. Hence $P_y$ is a high order polynomial of $E_x$ and $E_y$:

$$P_y(E_x, E_y) = P_{0y} + \epsilon_0 \left( \chi^{(1)}_{yx} E_x + \chi^{(1)}_{yy} E_y \right) +$$
$$\epsilon_0 \left( \chi^{(2)}_{yxx} E_x E_x + \chi^{(2)}_{yxy} E_x E_y + \chi^{(2)}_{yyx} E_y E_x + \chi^{(2)}_{yyy} E_y E_y \right) +$$
$$\epsilon_0 ( \chi^{(3)}_{yxxx} E_x E_x E_x + \chi^{(3)}_{yxxy} E_x E_x E_y + \chi^{(3)}_{yxyx} E_x E_y E_x +$$
$$\chi^{(3)}_{yxyy} E_x E_y E_y + \chi^{(3)}_{yyxx} E_y E_x E_x + \chi^{(3)}_{yyxy} E_y E_x E_y +$$
$$\chi^{(3)}_{yyyx} E_y E_y E_x + \chi^{(3)}_{yyyy} E_y E_y E_y) \quad (2)$$

After changing $E_x$ and $E_y$ and plotting $P_y$, the susceptibility tensor components are extracted using partial derivatives of $P_y$ with respect to $E_x$ and $E_y$. However to further simplify and speed up the calculations this multi-dimensional surface is cut along its axis by once assuming ($E_x=0$ and $E_y\neq 0$) which yields:

$$P_y(0, E_y) = P_{0y} + \epsilon_0 \left( \chi^{(1)}_{yy} E_y \right) + \epsilon_0 \left( \chi^{(2)}_{yyy} E_y E_y \right) +$$
$$\epsilon_0 \{ \chi^{(3)}_{yyyy} E_y E_y E_y \} \quad (3)$$

and in another run it is assumed that ($E_y=0$ and $E_x\neq 0$) i.e.:

$$P_y(E_x, 0) = P_{0y} + \epsilon_0 \left( \chi^{(1)}_{yx} E_x \right) + \epsilon_0 \left( \chi^{(2)}_{yxx} E_x E_x \right) +$$
$$\epsilon_0 \{ \chi^{(3)}_{yxxx} E_x E_x E_x \} \quad (4)$$

The other components of P i.e. $P_x$ and $P_z$ are written in a similar fashion. Fitting the Taylor expansions like equations (3) and (4) to polynomials is then a straightforward step in MATLAB® from which some tensor components of $\chi^{(2)}$ and $\chi^{(3)}$ are extracted [11].

**3 Results and Discussions** Table 1. and Table 2. list the diagonal and off-diagonal components of $\chi^{(2)}$ and $\chi^{(3)}$, respectively. The units are easily convertible to electrostatic unit (esu) by recalling that for $\chi^{(2)}$, 1 esu =

4.192×10$^{-4}$ m/V and for $\chi^{(3)}$, 1 esu = 1.398×10$^{-8}$ m$^2$/V$^2$. It is observed that the 2$^{nd}$ order nonlinear susceptibility is large as opposed to centro-symmetric bulk silicon. The *xxx* components of $\chi^{(2)}$ are 14.7, 133.5 and 141.4 pm/V for [110], [100], and [111] SiNWs, respectively. These values are still smaller than the experimentally measured $\chi^{(2)}$ = 600 pm/V in [5], however they are close to the values for silicon waveguides which are under tensile strain due to SiO$_2$ substrate i.e. $\chi^{(2)}$ = 60 pm/V. It should be mentioned that the precise value of $\chi^{(2)}$ reported in these experiments strongly depend on the measurement method. In electro-optically measured $\chi^{(2)}$ which is based on measuring the refractive index, the change of the latter due to accumulation or depletion of change carriers under DC field must be taken into account. This effect which is called Electric Field Induced Second Harmonic (EFISH) generation is under intensive research [2]. The enhanced value of $\chi^{(2)}$ for SiNWs presented here, is attributed to the sudden surface termination of SiNWs to Hydrogen atoms and the residual stress. This in turn emanates from canting of silicon dihydride units, SiH$_2$, after energy minimization. Whether the DFT method implemented in SIESTA® can capture the EFISH effect and the role of doping can be included in these simulations merit more study.

The 3$^{rd}$ order susceptibility values for *xxxx* and *yyyy* components are very close for each nanowire. For [110] and [100] they are one order of magnitude larger than that of bulk silicon. For [111] nanowire there is 100 times enhancement compared to bulk silicon. This can be directly interpreted as enhancement of Two Photon Absorption (TPA) coefficient, since the TPA process is of 3$^{rd}$ order and it can be related to $\chi^{(3)}$ using:

$$\beta_2(\omega) = \frac{\omega}{2n^2c^2\epsilon_0}\chi^{(3)}_{imag} \qquad (5)$$

Where n, c, and ω are refractive index, velocity of light in vacuum and frequency of photon, respectively. To estimate the TPA coefficient ($\beta_2$), we assume a photon of 2eV energy and a ballpark value of $\chi^{(3)}$ = 1×10$^{-18}$ m$^2$/V$^2$ = 7.15×10$^{-11}$ esu with a refractive index of n = 3. This results in $\beta_2$ ≈ 22 cm/GW which is very close to the values calculated for cubic silicon quantum dots [12]. However the method chosen in Iitaka et al. [12] is based on perturbation theory-based definition of $\chi^{(3)}$ which involves summations over many intermediate states and calculation of transition dipole matrix elements. This renders the method very time consuming as opposed to DFT-based method adopted here.

This estimation suggests prospects of enhancing TPA coefficient for SiNWs. Enhancing TPA is of importance for bandwidth broadening and increasing the efficiency of solar cells based on silicon nanowire arrays. Combination of this effect and modulation of linear absorption coefficient [13, 14] and light emission [15] due to mechanical strain promise applications of silicon nanowires and quantum dots for solar energy harvesting.

**Table 1** Diagonal elements of the 2$^{nd}$ and the 3$^{rd}$ order susceptibility tensors.

| Diagonal | | [110] | [100] | [111] |
|---|---|---|---|---|
| $\chi^{(2)}$ (pm/V) | *xxx* | 14.7 | 133.5 | 141.4 |
| | *yyy* | 43.3 | 21.5 | 272.3 |
| $\chi^{(3)}$ × 10$^{-18}$ (m$^2$/V$^2$) | *xxxx* | 1.4 | 2.0 | 12.88 |
| | *yyyy* | 1.6 | 1.8 | 12.91 |

The *xxx* and *yyy* components of $\chi^{(2)}$ are different from each other, however *xxxx* and *yyyy* components of $\chi^{(3)}$ are similar for each nanowire (See Table 1.). This further proves that the centro-symmetricity of each nanowire within cross sectional plane (*xy*) is broken and led to enhanced $\chi^{(2)}$ i.e. nonzero terms for E$^2$ within the Taylor expansions.

**Table 2** Off-diagonal elements of the 2$^{nd}$ and the 3$^{rd}$ order susceptibility tensors.

| Off-diagonal | | [110] | [100] | [111] |
|---|---|---|---|---|
| $\chi^{(2)}$ (pm/V) | *xyy* | 2.0 | 31.4 | 267.0 |
| | *yxx* | 81.1 | 225.6 | 81.0 |
| | *zxx* | 0.0027 | 0* | 0.0524 |
| | *zyy* | 0.0016 | 0* | 0.735 |
| $\chi^{(3)}$ × 10$^{-18}$ (m$^2$/V$^2$) | *xyyy* | 1.43 | 2 | 0.314 |
| | *yxxx* | 0.156 | 0.074 | 0.079 |
| | *zxxx* | 0.000033 | 0* | 0.00084 |
| | *zyyy* | 0.000057 | 0* | 0.0037 |

*These values are smaller than 10$^{-5}$

Table 2. shows the off-diagonal components of $\chi^{(2)}$ and $\chi^{(3)}$ for [110], [100] and [111] SiNWs. In this case the 2$^{nd}$ order components (*xyy*, *yxx*) are in the same order of magnitude as diagonal components in Table. 1. For off-diagonal terms a two order of magnitude drop for $\chi^{(3)}$ elements is observed. Interestingly this happens for those nanowires in which $\chi^{(2)}$ is still large i.e. [111]. This is again the manifestation of reduction of odd symmetry in **P** vs. **E,** hence a smaller $\chi^{(3)}$ value is extracted. As calculated P$_z$ values are at least four orders of magnitude smaller than P$_x$ and P$_y$, the extracted components like *zxx*, *zyy*, *zxxx* and *zyyy* are negligibly small.

After reviewing the computational results, a brief discussion about the shortcomings of the adopted DFT-based method is called for. As it was mentioned before in order to be considered as a perturbation, the external electric field must be applied normal to the longitudinal direction of the SiNWs (*z*). To further simplify the calculations, the two dimensional planes e.g. P$_y$(E$_x$, E$_y$) were cut along E$_x$ and E$_y$ axis. This means that some of the tensor components are missing which could be found using two dimen-sional slopes (gradients) of surfaces like $P_y(E_x,E_y)$ for example:

$$\chi^{(2)}_{yyx} = \frac{\partial^2 P_y(E_x,E_y)}{\partial y \partial x}, \quad \chi^{(3)}_{yxxy} = \frac{\partial^3 P_y(E_x,E_y)}{\partial^2 x \partial y} \quad (6)$$

Needless to say that the frequency dependency of nonlinear susceptibilities is not captured in this method. It is assumed that the frequency of light and its second and third harmonics are smaller than the bandgap ($E_g$) of nanowires, so there in no electronic excitation. On the other hand these frequencies must be higher than the vibrational frequencies of the nucleus [11]. Furthermore, high external electric field may lead to distortion of the unit cell and as a result of that a nonzero $E_z$. Although the orthonormalization procedure implemented in SIESTA® remedies the problem, however this slows down the convergence of energy minimizing process for higher electric fields.

**3 Conclusions** In summary it was shown how the nonlinear optical susceptibilities of silicon nanowires can be calculated using time independent linear scale DFT method implemented in SIESTA®. By applying an external electric field in allowed directions, calculating dipole polarization values, and fitting the results to Taylor expansion of P vs. E (equation 1), some of the components of susceptibility tensors can be extracted. It was observed that $\chi^{(2)}$ is enhanced up to 200 pm/V in close agreement with experiments with strained silicon waveguides. The $\chi^{(3)}$ values of SiNWs are comparable or better than that of bulk silicon by two orders of magnitude. Based on this an estimated value for two photon absorption coefficient was extracted which agrees well with those of silicon quantum dots. This study suggest possibility of bringing SiNWs into realm of nonlinear optics and benefitting from enhanced 2nd order nonlinearity due to surface termination, interfacing, and mechanical strain.

**Acknowledgements** I acknowledge access to the SHARCNET® supercomputing facilities of Ontario while I was working at University of Waterloo, Canada.